\documentclass[final,5p,times,twocolumn]{elsarticle}

\usepackage{amssymb}

\usepackage{graphicx,amsmath}
\usepackage{longtable}
\usepackage{booktabs}
\usepackage[hyphens]{url}
\usepackage[colorlinks=true,linkcolor=black, citecolor=blue, urlcolor=blue]{hyperref}

\usepackage{booktabs}	
\usepackage{float}

\usepackage{soul}
\usepackage{tabularx}
\usepackage{ltablex}
\usepackage[usenames, dvipsnames]{xcolor}
\usepackage{braket}
\usepackage{footnote}
\usepackage{xr}

\biboptions{authoryear,round}

\usepackage{lineno}

\begin{document}

\begin{frontmatter}


\title{Early identification of important patents through network centrality}

\author[label4,label1]{Manuel Sebastian Mariani}
\ead{manuel.mariani@unifr.ch}
 
\author[label4,label3,label1]{Mat{\'u}{\v{s}} Medo}
  
\author[label2,label5]{Fran\c{c}ois Lafond}

\address[label4]{Institute of Fundamental and Frontier Sciences, University of Electronic Science and Technology of China, Chengdu 610054, PR China}
\address[label1]{Department of Physics, University of Fribourg, 1700 Fribourg, Switzerland}
\address[label3]{Department of Radiation Oncology, Inselspital, Bern University Hospital and University of Bern, 3010 Bern, Switzerland}
\address[label2]{Institute for New Economic Thinking at the Oxford Martin School, University of Oxford, Oxford OX2 6ED, UK\\}
\address[label5]{Smith School for Enterprise and the Environment, University of Oxford, Oxford OX1 3QY\\}



\begin{abstract}
One of the most challenging problems in technological forecasting is to identify as early as possible those technologies that have the potential to lead to radical changes in our society.
In this paper, we use the US patent citation network (1926-2010) to test our ability to early identify a list of historically significant patents through citation network analysis. We show that in order to effectively uncover these patents shortly after they are issued, we need to go beyond raw citation counts and take into account both the citation network topology and temporal information. 
In particular, an age-normalized measure of patent centrality, called rescaled PageRank, allows us to identify the significant patents earlier than citation count and PageRank score. 
In addition, we find that while high-impact patents tend to rely on other high-impact patents in a similar way as scientific papers, the patents' citation dynamics is significantly slower than that of papers, which makes the early identification of significant patents more challenging than that of significant papers.
\end{abstract}


\end{frontmatter}

\newpage


\section{Introduction}
\label{intro}
While many inventions are granted a patent, only a small fraction of them represent ``important'' technological advances or will have a significant impact on the market.
As a result, a key problem in technological forecasting is to detect which patents are important as early as possible. The literature has designed various indicators of patent importance based on patent data analysis, and it has been found quite consistently (see Section \ref{sec:related}) that at least on average, important patents tend to receive more citations.  However, this relationship is typically noisy, which suggests that more sophisticated metrics could outperform simple citation count in identifying important patents. Importantly, it takes time for a patent to accumulate citations, which implies that simply counting the number of citations received by a patent may be effective for uncovering old important patents, but not to detect important patents shortly after they are granted. 

In this paper, we propose a network-based metric that identifies important patents better and earlier than citation count. Our metric, time-rescaled PageRank, was introduced by \citet{mariani2016identification} to identify expert-selected important papers in physics. It is built on Google's PageRank algorithm \citep{brin1998anatomy} by requiring that node score is not biased by node age. This metric is computationally efficient and thus can be applied on very large datasets~\citep{vaccario2017quantifying}.
Here we validate this metric on the US patent citation network (1926-2010), by evaluating its ability to detect the expert-selected ``important'' patents from \citet{strumsky2015identifying}.

We find that Google's PageRank outperforms raw citation count in identifying the important patents, which supports the idea that important patents tend to be cited by other important patents. 
This idea is further supported by the strong assortative degree-degree correlations observed in the network (Fig. \ref{degree_correlations} below): highly-cited patents are typically cited by other highly-cited patents significantly more than what we would expect by chance; at the same time, highly-cited patents tend to cite other highly-cited patents.
However, both PageRank and citation count are biased towards old patents; removing this bias is crucial to  compare young and old patents on the same scale~\citep{mariani2016identification}.

To demonstrate the usefulness of removing the age bias
in the context of technological forecasting, we evaluate the metrics' performance in identifying the important patents shortly after they are issued, and find that time-rescaled PageRank significantly outperforms citation count and original PageRank in the first $10$ years after issuing, approximately.
Finally, we use a time-respecting network null model \citep{ren2017time} to generate randomized networks where the individual patents' citation count dynamics is the same as in the original network. We find that 
both the observed degree-degree correlations and the performance advantage of PageRank-related metrics over citation count cannot be found in the randomized networks, which indicates that these properties emerge as a result of network effects that go beyond patents' neighborhood.

Our findings demonstrate that in order to timely identify the significant patents, both network topology and time information play an essential role. In more general terms, the advantage of PageRank-related metrics over citation counting metrics, together with the strong degree-degree correlations of the patent citation network, support the hypothesis that significant technological advances ``lean on the shoulders of giants'' in a similar way as scientific advances \citep{bornmann2010scientific}. 
Yet, we find that the citation dynamics of scientific papers and patents are characterized by substantially different timescales. As a result, because patents (on average) take more time than papers to accumulate citations, the early identification of significant patents is more challenging than that of significant papers.

\section{Related work}
\label{sec:related}
Broadly speaking, our work is related to those studying the relation between popularity metrics and significance in creative works such as scientific papers \citep{mariani2016identification,comins2017citation}, movies \citep{spitz2014measuring,wasserman2015cross,ren2017time}, or music albums \citep{monechi2017significance}.

In the context of patent analysis, it is well known that patents are of extremely different quality \citep{silverberg2007size}. While a direct measure of patent value is unavailable, patent data are very rich and there have been many attempts at providing indicators of patent value or novelty based on data contained in patent documents, such as the number of claims, the number and type of technology categories, the size of the patent family, and renewal fees, to give just major examples. By far the most widely used patent impact indicator is the number of citations received, and many studies have established a correlation between patent citations and patent value. For instance, \citet{trajtenberg1990penny}  found that to understand the evolution of the social value generated by the CT scan industry, it was better to count citations received by patents rather than simply counting patents.
\citet{albert1991direct} asked experts to rate the technical impact of patents in the area of silver halide technology, and found that highly cited patents received higher ratings. \citet{harhoff1999citation}, \citet{jaffe2000knowledge} and \citet{harhoff2003citations}, using survey data, found that citations were correlated with the value reported by the inventors.
\citet{lanjouw2004patent} collected several indicators of patent quality and concluded that citations and the number of claims were the most important indicators of quality. 
Recently, \citet{zhang2017entropy} proposed to weight $11$ indicators of patent value using the Shannon entropy, and selected forward citations as one of the most important indicators for technological value.
\citet{hall2005market} found that firm market value (Tobin's Q ratio) was correlated to the citation-weighted patent portfolio of the firms. 
\cite{carpenter1981citation} and \citet{fontana2013reassessing} compared patents associated with inventions that received a prize and patents from a control group, finding again evidence that ``important'' patents are more cited (the mean number of citations received was found to be about 50\% higher for important patents).

But in spite of the repeated evidence of the positive relationship between citations received and different indicators of value or quality, it is often
acknowledged that this relationship is very noisy \citep{harhoff1999citation}, thus leaving open the possibility that more elaborated indicators could outperform simple citations count in predicting patent value. Here we address two basic (and well-known) problems of evaluation by simply counting citations: when evaluating a given patent's score, it fails to take into account the importance of the citing patents \citep{narin1976evaluative}; and it fails to correct for the fact that young but potentially important patents did not have the time to accumulate a high number of citations.

The basic motivation for using citations received as an indicator of quality is that citations indicate some form of knowledge spillovers. As argued by \citet{jaffe2000meaning}, citations reflect the fact that either a new technology builds on an existing one, or that they serve a similar purpose. As a result, chains of citations allow us to trace the technological evolution, and hence patent centrality in the citation network can be used to score the patents. But not all measures of centrality are appropriate. For instance, in the case of patents, we want to value positively how many citations are received, but not necessarily how many citations are made. 

Whether the references made by a given patent can be used to infer the patent's importance is a delicate issue.
In principle, one could argue that a patent with many references has high potential, because it draws from many existing inventions. But an opposite argument could be made as well, because a patent with many references makes it also (legally) clear that its claims are somewhat limited by the claims of the cited patents -- in that sense, references indicate a limitation of novelty. It is not yet well-understood which of these two arguments is the most appropriate, and the empirical evidence so far is inconclusive \citep{jaffe2017patent}; here, we will consider that citations received are a weaker signal of importance when they come from patents that make a lot of references”.

Based upon the aforementioned considerations, Google's PageRank centrality \citep{brin1998anatomy} is especially suited for identifying important patents for three reasons: (i) It takes into account how many citations are received by a patent, (ii) It takes into account how many citations are received by the citing patents, and (iii) it takes into account that citations from patents that have many references are less indicative of the cited patent's quality.

We are not the first ones to suggest that PageRank \citep{lukach2007ranking,bedau2011evidence,shaffer2011entrepreneurial,dechezlepretre2014knowledge,bruck2016recognition} and similar eigenvector-based metrics \citep{corredoira2015measuring} can be computed on patent citation networks to identify important patents.  However, robust evidence that PageRank is more effective than citation count in identifying the key patents is still lacking. In addition, both citation counts and PageRank fail to take into account the dynamic, evolving nature of the citation network. Because the patent system grows with time, older patents tend to have more citations simply because they have been there for a longer time and, on top of that, the preferential attachment mechanism \citep{valverde2007topology} further magnified their advantage. This problem has been long acknowledged and the usual solution is either to limit citation counts to a fixed citation “time span”, such as the first five years after issuing (e.g., \citet{lanjouw2004patent}), or to control for the grant year in regressions (e.g., \citet{kogan2012technological}).

Here, we propose an alternative approach, put forward recently by \citet{mariani2016identification} in the context of scientific publications, which can be applied equally well to citation counts and other centrality metrics, and produces a single score without (or with dramatically reduced) age bias.

Our work complements other efforts to identify important items using citation networks. For instance, Comins and Leydesdorff (2017) report that Reference Publication Year Spectroscopy, a method that looks at the temporal distribution of cited references, is able to identify the biomedical research milestones listed by experts. In the patent literature, \citet{castaldi2015related} proposed to identify ``superstar" patents as those in the extreme right tail of the citation count distribution, where a power law behavior was observed. Another popular approach, main path analysis, was introduced in the bibliometric literature by \citet{hummon1989connectivity}, and further developed and applied to patents by \citet{verspagen2007mapping}. In the spirit of the betweenness centrality, it seeks to extract important nodes and edges based on how often geodesic paths pass through them, thus revealing continuity or disruption in technological trajectories. This aspect was exemplified by \citet{martinelli2012emerging} for the telecommunication switching industry, and by \citet{epicoco2013knowledge} for the green chemistry sector. \citet{triulzi2017predicting} measured patent centrality using a normalized version of centrality metrics, and found that technological domains with central patents also tend to have faster technological improvement rates (a separately measured indicator of progress in technological performance). Finally, as a last example of this rich literature, \citet{martinelli2014measuring} proposed to measure knowledge persistence by giving higher value to patents that are cited by patents that don’t cite many patents \-- an idea that we will use here too, as PageRank normalizes the received citations by the outdegree of the citing nodes.

In this work, we focus on comparing PageRank with citation counts, and age-rescaled metrics with non-rescaled metrics. This allows us to evaluate whether network-based metrics outperform raw citations counts, and to determine over which range of time the rescaling procedure allows us to better identify the significant patents. In addition, because our analysis follows closely the study of milestone physics papers by \cite{mariani2016identification}, we are able to evaluate the similarities and differences between the scholarly and patent citation data. We find that patents take much longer than papers to receive citations, which makes it harder to identify important patents early on.

\begin{figure*}
\centering
\includegraphics[scale=0.6]{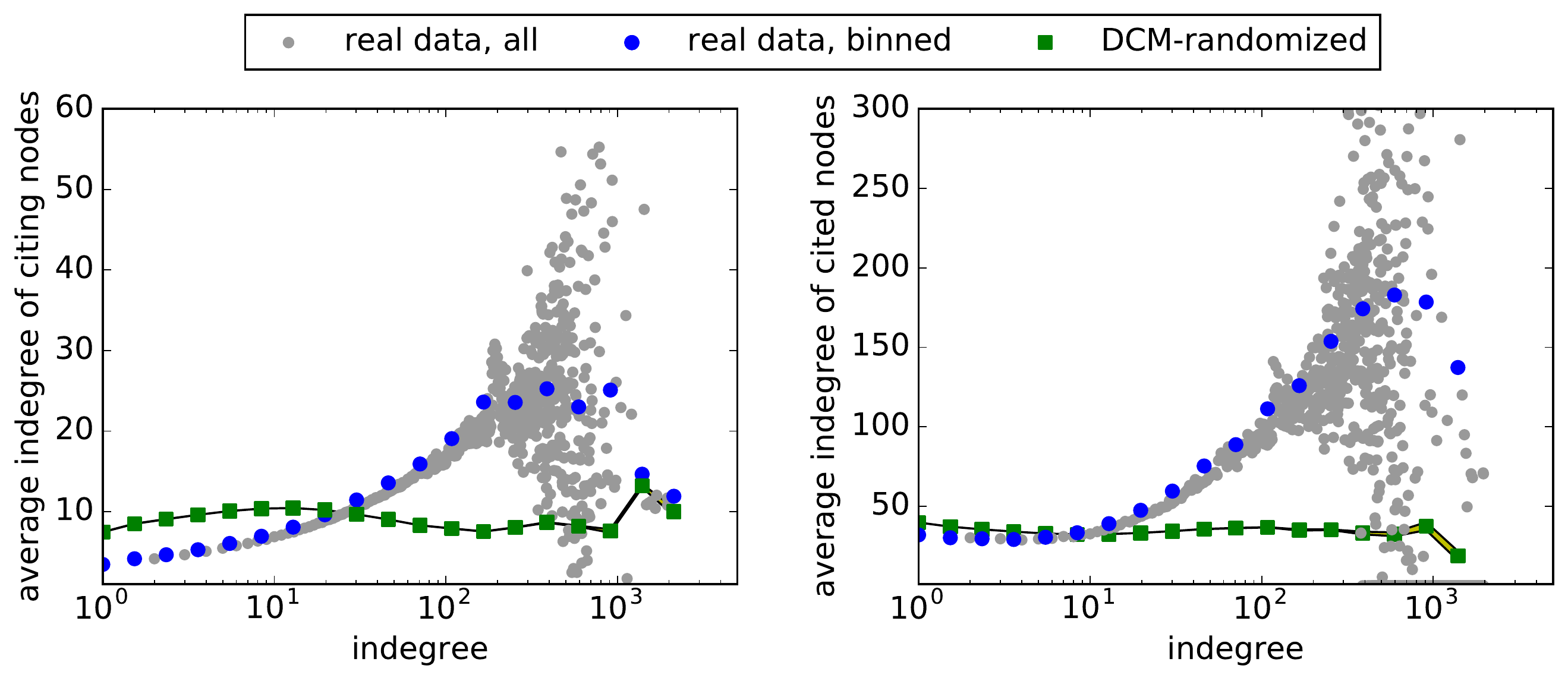}
\caption{Degree-degree correlations in the US patents' citation network. The gray circles represent the observed average neighbors' indegree for all the indegree values; the blue circles represent the same information in a histogram with equal bin length on logarithmic scale; the green squares represent the mean behavior observed within ten realizations of the dynamic configuration model (see Section \ref{sec:dcm} for a description of the model); the shadows around the line that connect the green squares represent one standard deviation around the mean. 
}
\label{degree_correlations}
\end{figure*}

\section{Data}

\subsection{The U.S. patent citation network}
We analyzed the US patents dataset collected by \cite{kogan2012technological} that spans the period between 01-01-1926 and 11-02-2010. As compared to the well-known NBER patent data, this dataset has a vastly improved coverage.
We pre-processed the data to only keep the citations between patents that were issued within this temporal period, removing thereby the citations to patents issued before 01-01-1926.
The resulting citation network analyzed in this paper is composed of $N=6,237,625$ nodes (patents) and $E=45,962,301$ directed edges (citations) between them. 

In this dataset, the in- and out-degree distribution of the US patent citation network are in agreement with previous findings \citep{valverde2007topology,csardi2007modeling,silverberg2007size}: the two distributions are relatively broad and span more than three orders of magnitude.

In previous works, \cite{mariani2016identification} found that PageRank-related metrics outperform citation-counting metrics in identifying significant nodes in a scientific paper citation network, whereas the same does not happen in a movie citation network \citep{ren2017time}.
Additionally, \cite{ren2017time} found remarkably different degree-degree correlations for the two networks: the papers' citation network is strongly assortative, whereas the movies' citation network is basically uncorrelated.
This observation led \cite{ren2017time} to suggest that the relative performance of PageRank-related and citation-counting metrics may be related to the network correlation patterns: when the network is uncorrelated, PageRank and indegree bring basically the same information \citep{fortunato2008approximating}; when there are significant structural correlations, PageRank brings additional information that may be valuable to improve ranking performance.

Fig.~\ref{degree_correlations} shows that the US patent network exhibits strong degree-degree correlations\footnote{The assortativity plot used here is arguably the simplest method to visualize network structural correlations, as simply represents the average (in- or out-)degree of nodes' neighbors as a function of node (in- or out-)degree. Since node centrality is related to incoming connections, we focus here on the average indegree of citing and cited nodes as a function of node indegree.}: highly-cited patents tend to be cited by other highly-cited patents, and to cite other highly-cited patents. This assortative pattern cannot be explained by a null model that preserves the temporal evolution of node degree (\citep{ren2017time}, see Section \ref{sec:dcm} for details), which suggests that it is a genuine network effect.

In agreement with similar findings for scientific papers \citep{ren2017time,bornmann2010scientific}, Fig.~\ref{degree_correlations}A suggests that high-impact patents are able to inspire other high-impact patents more than expected by chance, whereas low-impact patents tend to be cited by other low-impact patents; at the same time (Fig. \ref{degree_correlations}B), high-impact patents rely on other high-impact patents more heavily than expected by chance.
Following \cite{ren2017time}, the strong degree-degree correlations in the patent citation network opens the door to the possibility that metrics that take higher-order network effects into account outperform simple citation counts.

\subsection{Expert-selected historically significant patents}
\label{sec:strumsky}
In a recent work, \citet{strumsky2015identifying} listed $175$ patents carefully selected ``on the basis of consultation with engineers, scientists, historians, textbooks, magazine articles, and internet searches''. The patents in the list ``all started technological pathways which affected society, individuals and the economy in a historically significant manner'' \citep{strumsky2015identifying}. These significant patents thus provide a good ``ground-truth'' set of patents that can be used to discern the ability of different metrics to uncover the significant patents. The complete list of significant patents can be found in Appendix C of \cite{strumsky2015identifying}; the list is quite heterogeneous and comprises patents ranging from simple objects 
that are part of our everyday life (like the toothbrush and the post-it note)
to more sophisticated inventions (like the Game Boy and the Desk Jet printer).

Presence in the list of significant patents by Strumsky and Lobo is a binary variable: a patent is either in the list or not; we can therefore study the ability of the metrics to rank these outstanding patents as high as possible, in agreement with the main goals of this paper. 
While there are $175$ significant patents in the Strumsky-Lobo list, we restrict our analysis to those patents that were issued within our dataset's temporal span, and remove the design patents which are absent in our dataset. This leaves us with $M_0=112$ significant patents.

\begin{table*}[t]
\centering
\begin{tabular}{lp{5cm}p{5cm}}
\toprule
  & Static & Age-rescaled \\
\midrule
Citation-counting &
\textbf{Citation count}, $c$\newline
A patent is important if it is cited by many other patents &
\textbf{Rescaled citation count}, $R(c)$\newline
Built on citation count by requiring that patent score is not biased by node age\\[4pt]
Network-based &
\textbf{PageRank score}, $p$\newline
A patent is important if it is cited by other important patents &
\textbf{Rescaled PageRank score}, $R(p)$\newline
Built on PageRank score by requiring that patent score is not biased by node age\\
\bottomrule
\end{tabular}
\caption{Metrics considered in this paper together and their main assumptions.}
\label{table:metrics}
\end{table*}

\section{Methods}
In this section, we define the metrics used to identify important patents, and the indicators of performance that we use to evaluate them.
Many network centrality metrics \citep{lu2016vital,liao2017ranking} and bibliometric indicators \citep{waltman2016review} have been devised in the literature. Here, we narrow our focus to four metrics (see Table \ref{table:metrics} for a summary): citation count $c$, PageRank score $p$, (age-)rescaled citation count $R(c)$ and (age-)rescaled PageRank $R(p)$. Differently from citation count, PageRank score takes the whole network structure into account and weights citations differently according to the centrality of the citing nodes. Rescaled citation count and rescaled PageRank score are obtained from citation count and PageRank score, by explicitly requiring that node score is not biased by node age (see details below).

\begin{figure}
\centering
\includegraphics[width=7.5cm]{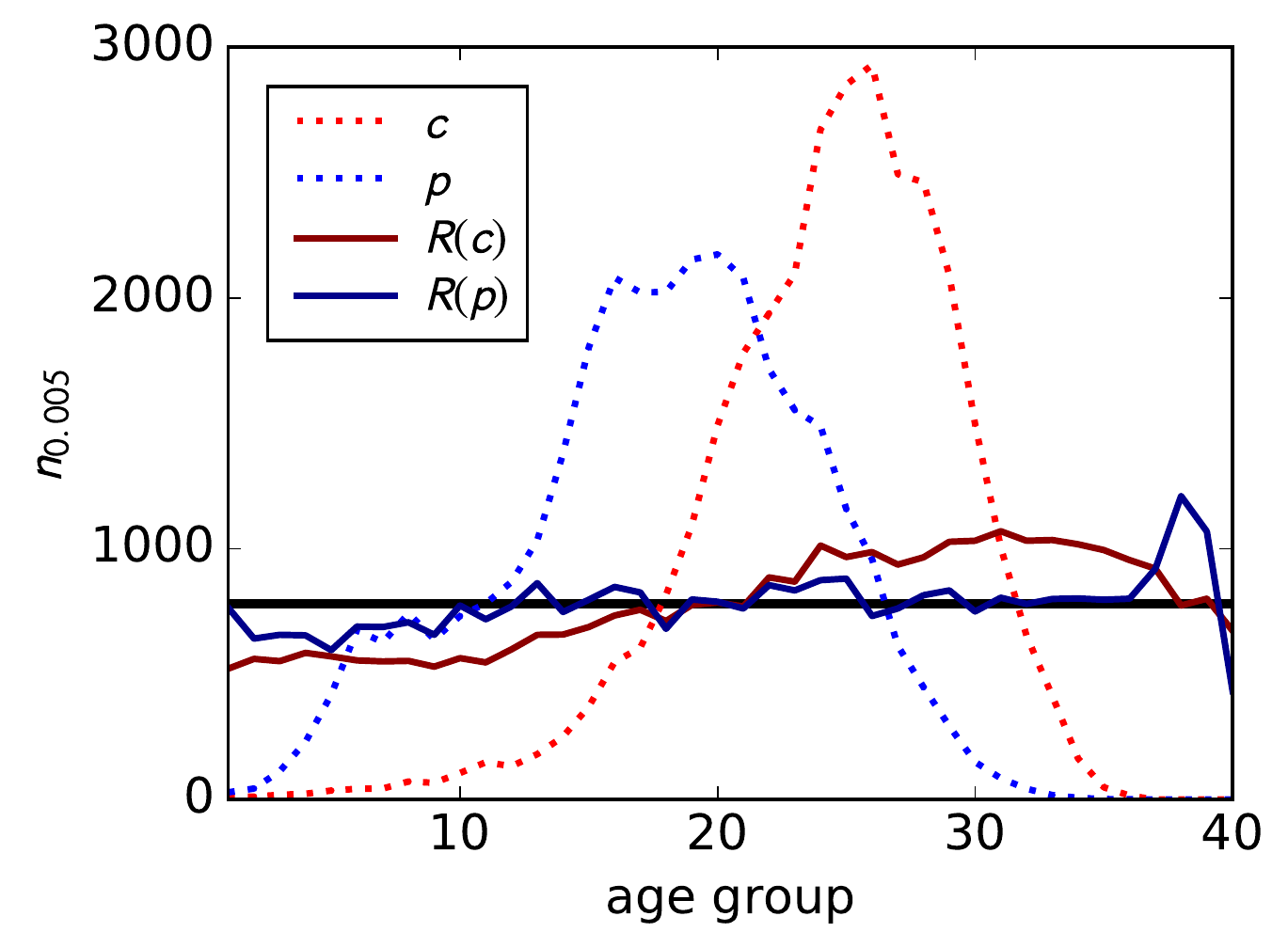}
\caption{Bias by node age of the rankings by the metrics. Patents are divided by their age into 40 equally-sized groups; the bias by patent age is represented by the number $n_{0.005}$ of patents from each age group in the top $f=0.5\%$ of the overall patent ranking. The black horizontal line represents the expected unbiased value $n^{(0)}_{0.005}=0.005\,N/40$. Results for different (small) values of $f$ are qualitatively similar.}
\label{bias}
\end{figure}

\subsection{Static patent-level metrics}
\label{sec:1}

\paragraph{Citation count, c}
The citation count of a given patent is simply the total number of citations the patent has received so far. In terms of the patent citation network's adjacency matrix $\mathsf{A}$ ($A_{ij}$ is equal to one if patent $j$ cites patent $i$, zero otherwise), the citation count $c_i$ of patent $i$ is defined as $c_i=\sum_{j}A_{ij}$; $c_i$ is referred to as the node $i$'s indegree in the network science language \citep{newman2010networks}. Ranking the patents by citation count assumes that \emph{a patent is important if it is cited by many other patents}. 

The ranking by citation count is strongly biased by node age in our dataset. To visualize and quantify this bias, we divide the $N$ patents into $40$ equally-sized age-groups based on age.
We then count how many patents from each age group are in the top-$f$ fraction of the patent ranking by $c$. For an ideal unbiased ranking, for each age group, we would expect $n_{f}^{(0)}=f\,N/40$ patents in the top-$f$ fraction, with small deviations.
The result is strikingly different for citation count (see Fig.~\ref{bias}) which underestimates both the oldest and the most recent patents in the dataset. 
While the bias against recent patents is expected (they have had less time to accumulate citations), the bias against older patents is more surprising, and it can be due to a variety of factors such a higher propensity to cite patents available electronically, a prevalence of patents in technological domains for which less citations tend to be made, patent citation patterns changing with time -- for a discussion of these and other reasons for citation bias, see \citet{jaffe2017patent}. 
To counteract this bias, we use a simple normalization procedure, described in paragraph~\ref{sec:rescaled}. 


\paragraph{PageRank, p}
Google's PageRank is a node ranking algorithm introduced by \citet{brin1998anatomy} with the original goal to rank websites in the World Wide Web. Since then, the algorithm has found applications in a broad range of real systems \citep{gleich2015pagerank,liao2017ranking}. 
The PageRank score $p_i$ of node $i$ is defined through the equation \citep{berkhin2005survey}
\begin{equation}
p_i=\alpha\,\sum_{j:k^{out}>0}\frac{A_{ij}}{k^{out}_j}\,p_j+\alpha\,\sum_{j:k^{out}=0}\frac{p_j}{N}+\frac{1-\alpha}{N}
\label{pr}
\end{equation}
where $k^{out}_j=\sum_{l}A_{lj}$ is the number of references made by patent $j$ ($k^{out}_i$ referred to as the node $i$'s \emph{outdegree} in the network science language) and the term $(1-\alpha)/N$ represents the so-called ``teleportation term'' \citep{berkhin2005survey,gleich2015pagerank}.
The algorithm is built on the thesis that \emph{a node is important if it is cited by other important nodes} \citep{franceschet2011pagerank}: the score of a given patent $i$ depends linearly on the scores of the patents that cited patent $i$. We set $\alpha=0.5$ which is the common choice in citation networks \citep{chen2007finding,walker2007ranking,bruck2016recognition}.

In practice, the vector of PageRank scores can be obtained from Eq.~\eqref{pr} by the power iteration method. Starting from a uniform score vector $p_i^{(0)}=1/N\,\forall i$, we iteratively update the scores according to the equation \citep{berkhin2005survey} 
\begin{equation}
p_i^{(n+1)}=\alpha\,\sum_{j:k^{out}>0}\frac{A_{ij}}{k^{out}_j}\,p_j^{(n)}+\alpha\,\sum_{j:k^{out}=0}\frac{p_j^{(n)}}{N}+\frac{1-\alpha}{N}.
\end{equation}
Note that the previous equation is the master equation of a two-fold stochastic process on the network where at each step $n$, a random walker either performs a jump along the network edges (with probability $\alpha$), or ``teleports'' to a randomly chosen node in the network (with probability $1-\alpha$).
The PageRank vector of scores $\mathbf{p}=\{p_i\}$ can therefore be interpreted as the stationary state of this Markov process.
We halt the iterations when 
\begin{equation}
\sum_i \big\lvert p^{(n)}_{i}-p^{(n-1)}_{i}\big\rvert < \epsilon,
\end{equation}
where we set $\epsilon=10^{-9}$. This procedure guarantees convergence after a number of iterations smaller than $\log\alpha/\log\epsilon$, independently of $N$ \citep{berkhin2005survey}.

While PageRank's basic premise is plausible, the algorithm is \emph{static}, whereas real networks evolve in time. This causes the ranking by the algorithm to be severely biased by node age in growing networks \citep{chen2007finding,mariani2015ranking,mariani2016identification,vaccario2017quantifying}.
The ranking by PageRank is strongly biased by node age also in our dataset (see Fig.~\ref{bias}). 
PageRank's bias in the patent citation network has different features with respect to its bias in papers' citation network reported by \cite{mariani2016identification}.
While in both datasets recent nodes are strongly disadvantaged by the algorithm, the oldest patents are not the most overvalued by the PageRank algorithm as opposed to what has been observed for papers \citep{mariani2016identification}.
This is a direct consequence of the significantly smaller citation count of the oldest patents. The peak of $n_{0.005}(p)$ is shifted to the left from the peak of $n_{0.005}(c)$, which means that PageRank nevertheless tends to favor older nodes with respect to citation count.

\subsection{Time rescaled metrics R(p) and R(c)}
\label{sec:rescaled}
The strong age bias of the rankings by citation count and PageRank score implies that patents that appeared in some time periods are much more likely to rank well than other patents, independently of their properties such as novelty and significance. In bibliometrics \citep{radicchi2008universality,waltman2016review} and patent analysis \citep{triulzi2017predicting}, it is common to attempt to suppress this bias by age through various normalization procedures.

Here, we apply the rescaling procedure proposed by \citet{mariani2016identification} to citation count and PageRank.
The rescaling procedure consists of comparing the score $s_i$ of a given patent $i$ with scores of the patents that belong to a reference-set of patents of similar age\footnote{A potential limitation of this approach is that by comparing each patent's score with only the scores of patents of similar age, it may underestimate the importance of patents that happened to be issued in periods during which many breakthrough inventions took place. However, despite the well-known theory of Kondratiev waves and innovation clustering, robust empirical evidence for the existence of such periods is weak and debated. For instance, \cite{silverberg2003breaking} found no evidence for innovation clustering in a list of basic inventions,whereas \citet{korotayev2011kondratieff} found an evidence of Kondratiev cycles in the world-level patent output per inhabitant.} as patent $i$. 
By labeling the patents in order of decreasing age\footnote{We order by increasing ID those patents that are issued on the same day.}, the reference set is the set of $\Delta$ patents $j$ such that $\lvert i-j\rvert < \Delta/2$.\footnote{The temporal window is defined in a slightly different way for patents close to the beginning and the end of the dataset. For the $\Delta/2$ patents closest to the beginning (end) of the dataset, the temporal window is given by the $\Delta$ oldest (most recent) patents in the dataset.} Constructing the set of comparable patents based on a continuously moving window centered on a focal patent is advantageous with respect to grouping the patents by year as the latter results in imposing a sharp distinction between patents granted very closely in time but on either side of the January 1st boundary.

Denoting with $\mu_i(s)$ and $\sigma_i(s)$ the mean value and the standard deviation, respectively, of score $s$ over the patent $i$'s reference set, the rescaled score $R_i(s)$ of patent $i$ is given by 
\begin{equation}
R_i(s)=\frac{s_i-\mu_i(s)}{\sigma_i(s)}.
\end{equation}
In this work, we set $\Delta=15,000$, yet our results are robust with respect to other choices of $\Delta$ (not shown here).

As shown in Fig.~\ref{bias}, the rescaled scores $R(c)$ and $R(p)$ are much less biased by node age than the original scores $c$ and $p$: $n_{0.005}(R(c))$ and $n_{0.005}(R(p))$ are remarkably stable across different age groups, and their value is always close to the expected unbiased value $n_{0.005}^{(0)}$. In agreement with \cite{mariani2016identification,vaccario2017quantifying,liao2017ranking}, this shows that the proposed rescaling procedure is effective in suppressing the temporal biases of the static metrics. 
By giving to old and recent patents the same chance of appearing to the top of the ranking, we expect the rescaled metrics to bring a substantial advantage in identifying valuable patents shortly after issuing. As the rankings by static metrics are biased toward old patents, we also expect the rescaled metrics' advantage in identifying significant patents to shrink (and eventually vanish) as we consider older significant patents. 
These hypotheses are validated in the next Section. 

\subsection{Evaluation of the metrics' performance in identifying the significant patents}

To make quantitative statements on the ability of the metrics to single out the significant patents of different age, we introduce two evaluation metrics: the average ranking ratio and the identification rate.

\paragraph{Average ranking ratio}
A straightforward way to assess the metrics' performance in identifying the significant patents would consist in calculating the average ranking position of the significant patents $t$ years after they are issued. However, this simple measure is highly sensitive to the ranking position of the lowest-ranked items \citep{mariani2016identification}. 

To prevent this shortcoming, it is preferable to measure the \emph{average ranking ratio} of the target items by the different metrics \citep{mariani2016identification}, which is defined as follows. Denoting the rank of patent $i$ by metric $m$ as $r_i(m)$,
the ranking ratio of metric $m$ is defined as $\hat{r}_{i}(m)=r_{i}(m)/\min_{m'}\{r_{i}(m')\}$.  The metric achieves the best-possible ranking ratio of one if it ranks a given significant patent best of all metrics; the lower the value, the better.
The \emph{average ranking ratio} $\braket{\hat{r}}(m)$ of metric $m$ is the average of the ranking ratios $\hat{r}_{i}(m)$ of all significant patents, and it quantifies how much the metric underperforms, on average, with respect to the best-performing metric. A metric that outperforms all the other metrics for all the target nodes achieves an average ranking ratio $\braket{\hat{r}}=1$; larger values of $\braket{\hat{r}}$ indicate a worse performance.

\paragraph{Identification rate} The identification rate $f_z(m)$ -- commonly referred to as \emph{recall} in the information filtering community \citep{lu2012recommender} -- of a given metric $m$ is defined as the fraction of significant patents that are ranked among the top $z\,N$ patents by metric $m$. Hence, while the average ranking ratio takes all significant patents and their ranking into account, the identification rate measure focuses on the top-items by each ranking.  

\subsection{Evaluating the evolution of the metrics' performance with patent age}
\label{sec:age_evaluation}
To uncover the metrics' ability to early identify the significant patents, 
we evaluate the metrics' average ranking ratio and identification rate as a function of patent age. In this way, we are able to untangle the role of patent age in determining the metrics' performance; for example, a metric that is biased toward old patents only performs well in detecting \emph{old} important patents, whereas we expect good early-identification metrics to perform well in detecting \emph{recent} important patents.


To untangle the role of patent age in determining the metrics' performance, we dissect the network evolution by computing the rankings by the metrics every six months. At each ranking computation time $t^{(c)}$, only the patents issued before $t^{(c)}$ are included in the analysis. For significant patent $i$ (issued at time $t_i$), we measure the age $\Delta t=t^{(c)}-t_i$ of the significant patent at time $t^{(c)}$. 
Then, we determine its ranking ratio values $\hat{r}_i(m;\Delta t)$ for all considered metrics $m$.
Patent $i$'s ranking ratio $\hat{r}_i(m;\Delta t)$ contributes to metric $m$'s average ranking ratio $\braket{\hat{r}}(m;\Delta t)$ for $\Delta t$ year-old patents.
After having analyzed the whole network history, we can thus determine the average ranking ratio $\braket{\hat{r}_i}(m;\Delta t)$ of metric $m$ for $\Delta t$ years old patents as the average of $\hat{r}_i(m;\Delta t)$ over all the significant patents included in the analysis.


In the same way, we define the identification rate $f_z(m;\Delta t)$ of metric $m$ for $\Delta t$ years old patents as the fraction of significant patents that were ranked among the top $z\,N$ patents by metric $m$ when they were $\Delta t$ years old. 

\section{Results}
\label{section:results}

\subsection{Metrics' performance on the time-aggregate network}
\label{sec:aggregate}
\begin{figure*}[t]
\centering
\includegraphics[width=15cm]{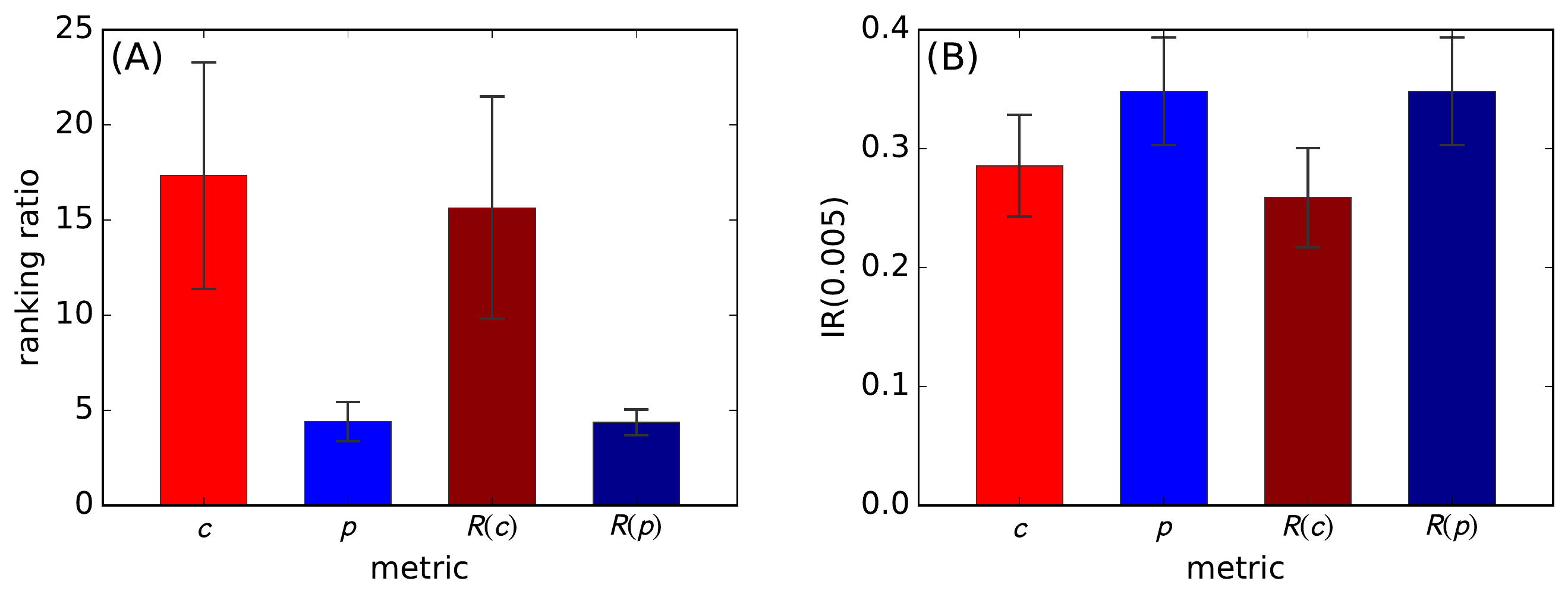}
\caption{Performance of the metrics in identifying the significant patents from the list by Strumsky and Lobo, as measured by the metrics' average ranking ratio (panel A, the lower the better) and their identification rate (panels B, the higher the better) evaluated on the complete patent citation dataset} 
\label{histo}
\end{figure*}

We start by assessing the average ranking ratio (Fig.~\ref{histo}A) and the identification rate (Fig. \ref{histo}B) of the metrics on the whole dataset.
The results show a clear advantage of the network-based metrics, $p$ and $R(p)$, over the citation-counting metrics. According to the average ranking ratio (Fig.~\ref{histo}A), rescaled PageRank is the best-performing metric with a small margin over PageRank and a large margin over raw and rescaled citation count. Rescaled PageRank and PageRank also achieve the highest identification rates (Fig.~\ref{histo}B). 

To understand where the gaps between the metrics stem from, we inspect the patents that give the largest contribution to $c$ and $R(c)$'s ranking ratio -- i.e., the patents that are ranked much better by $p$ and $R(p)$ than by $c$ and $R(c)$. We find a significant contribution coming from patent $4,237,224$ (``Process for producing biologically functional molecular chimeras'', $c=285$), which is ranked $2$nd by $R(p)$ ($\hat{r}=1$), $3$rd by $p$ ($\hat{r}=1.5$), $1079$th by $R(c)$ ($\hat{r}=539.5$), and $1181$st by $c$ ($\hat{r}=590.5$). 
Importantly, this patent gives the same contribution (equal to one) to all metrics' identification rate as all the metrics rank it among the top-$0.5\%$ patents. This example shows well that patents that are ranked at the top of the ranking by all metrics can have very different ranking ratio values.
The second largest contribution to $c$'s and $R(c)$'s average ranking ratio comes from patent $4,438,032$ (``Unique T-lymphocyte line and products derived therefrom'', $c=73$), which is ranked $253$rd by $p$ ($\hat{r}=1$), $562$nd by $R(p)$ ($\hat{r}=2.2$), $48,742$nd by $c$ ($\hat{r}=192.7$), $66,014$th by $R(c)$ ($\hat{r}=260.9$). 
To check that the advantage of network-based metrics was not entirely due to these two patents, we have excluded them from the analysis and recalculated the metrics' average ranking ratio. PageRank and rescaled PageRank remain the two best-performing metrics ($\braket{\hat{r}}(p)=4.2$, $\braket{\hat{r}}(R(p))=6.1$), yet their edge over the link-counting metrics ($\braket{\hat{r}}(c)=8.1$, $\braket{\hat{r}}(R(c))=10.0$) significantly decreased.

\subsection{Age-rescaling matters most for young patents}
\label{dynamics}
\begin{figure*}
\centering
\includegraphics[width=15cm]{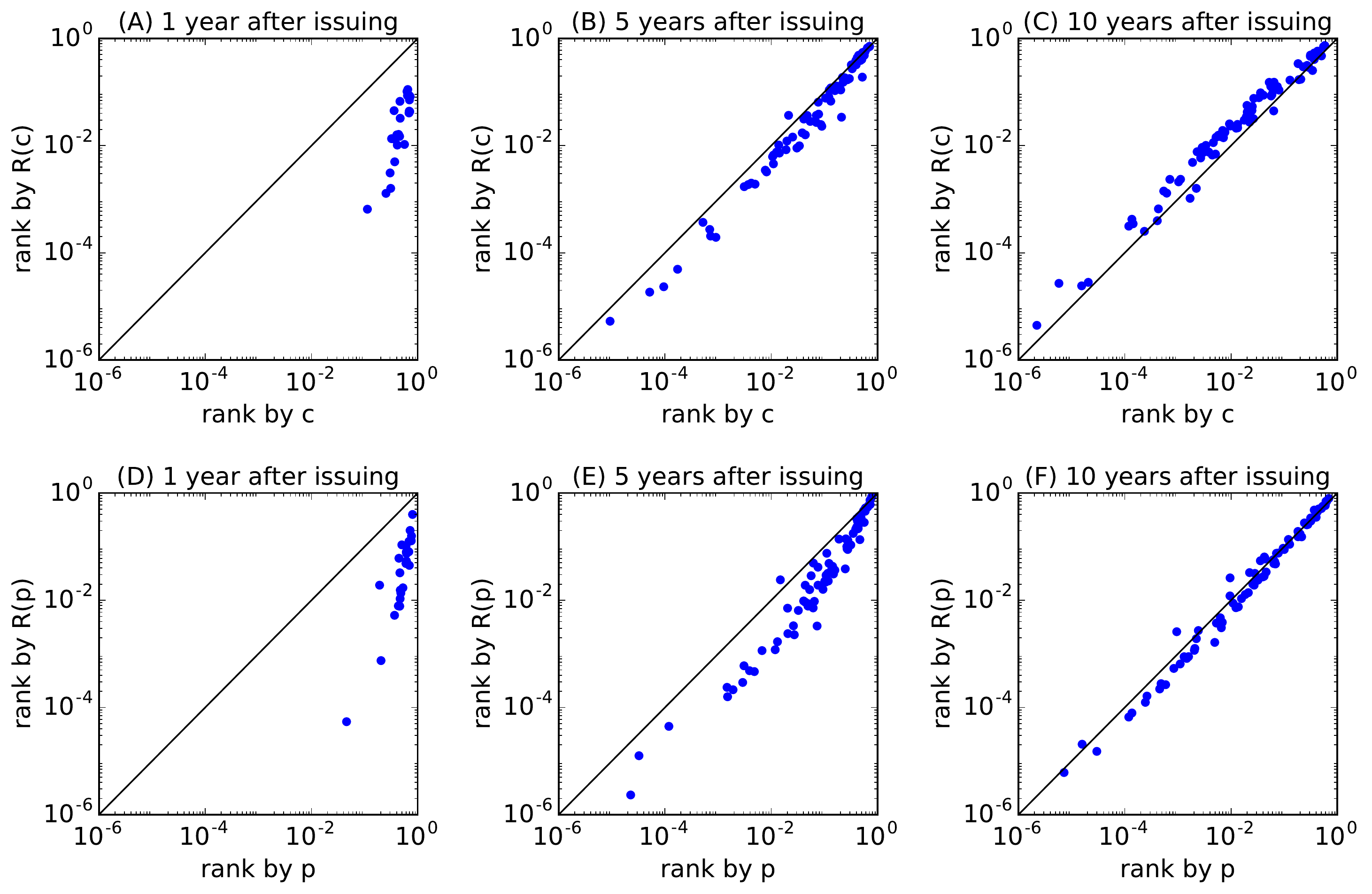}
\caption{A comparison of the relative rankings (the lower, the better) of the significant patents by $c$ and $R(c)$ one (panel A), five (panel B) and ten (panel C) years after issuing. Only the patents that received at least one citation at a given age are included. The same comparison between $p$ and $R(p)$ is shown in panels D--F.}
\label{rankings}
\end{figure*}

While the analysis of the previous Section reveals important differences among the metrics, the main goal of this manuscript is to reveal the dependence of the metrics' performance as a function of patent age, and to assess the metrics' ability to early-identify the significant patents. To this end, by following the procedure described in Section \ref{sec:age_evaluation}, we consider the ranking positions\footnote{The ranking positions considered in this paper are always normalized by the size of the system at the time when the ranking is computed.} of the group of expert-selected significant patent by \cite{strumsky2015identifying} one (Figs.~\ref{rankings}A,D), five (Figs.~\ref{rankings}B,E) and ten (Figs.~\ref{rankings}C,F) years after issuing. Due to their lack of time bias, the rescaled metrics rank the significant patents much better than the corresponding static metrics one year after issuing (see Figs.~\ref{rankings}A,D). On the other hand, the ranking positions by rescaled and static metrics are comparable ten years after issuing (see Figs.~\ref{rankings}C,F).

The evolution of the ranking position of the significant patents as evaluated by $p$ and $R(p)$ is shown in Supplementary Movie M1; the same for $c$ and $R(p)$ is shown in Supplementary Movie M2. The moving dots in these movies represent the significant patents, and the displacements of the dots represent the change in the significant patents' ranking position as they get older\footnote{We only represent the significant patents after they have received their first citation. This is the reason why during the dynamics, some dots appear on the plane out of nowhere.}. 
Movies M1 and M2 show that short after issuing, all significant patents are ranked higher by rescaled PageRank than by PageRank and citation count, respectively, consistently with Figs.~\ref{rankings}A,D. In movie M1 that compares the rankings by $p$ and $R(p)$, as the significant patents get older, the entity of their displacements in the ranking plane diminish, and they gradually drift toward the diagonal of the plane, which means that the gap between their ranking position by $p$ and $R(p)$ shrinks. 
After ten years, most of the significant patents lie close to the diagonal, which indicates that the rankings of the significant patents by $p$ and $R(p)$ are comparable.

\begin{figure*}
\centering
\includegraphics[width=18cm]{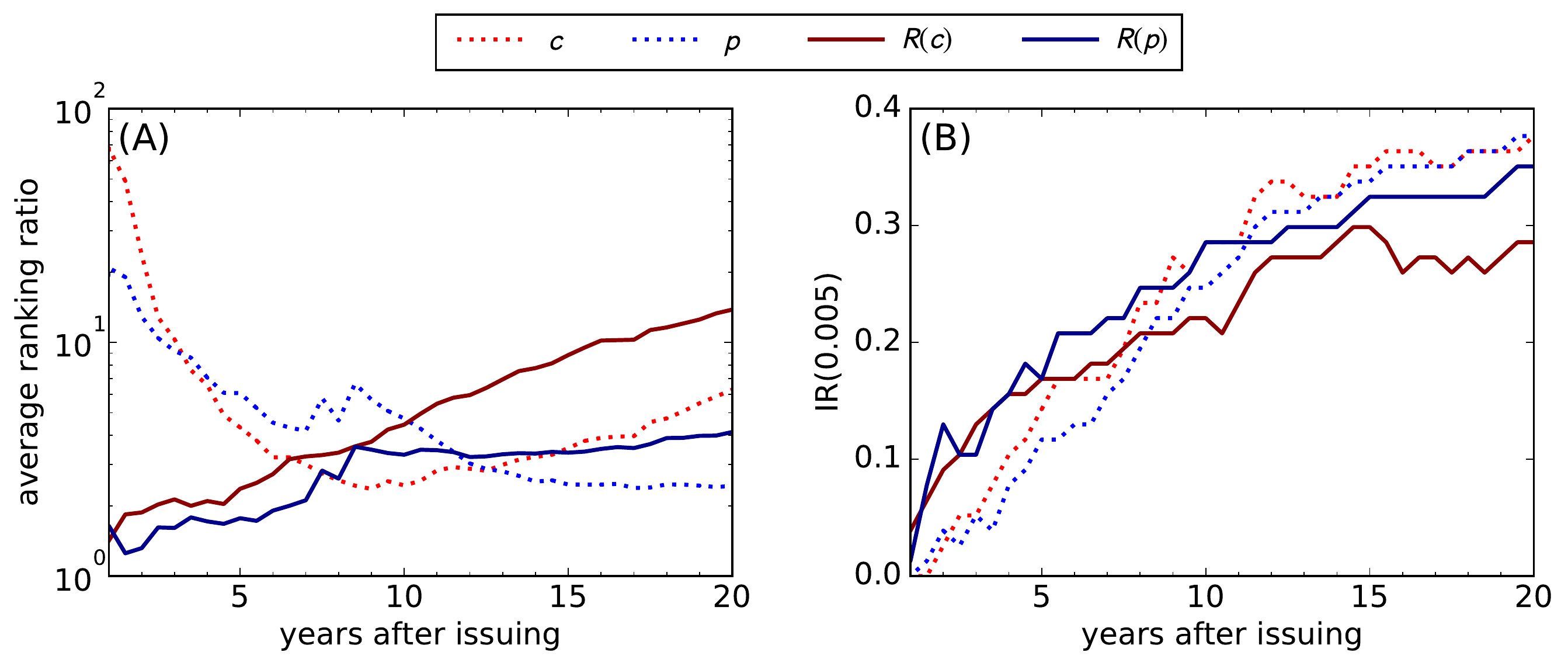}
\caption{Performance of the metrics in identifying the significant patents from the list by Strumsky and Lobo over a $20$-year time window after their issuing, as explained in the main text. \emph{(A)} Average ranking ratio as a function of patent age. \emph{(B)} Identification rate as a function of patent age.}
\label{short} 
\end{figure*}

\subsection{Comparison of the four metrics' performance for different patents' age}
The above-discussed Fig.~\ref{rankings} and Supplementary Movies M1--M2 show that the age of the significant patents has a large impact on the ability of the metrics to identify them.
The goal of this section is to quantify the magnitude and the duration of the advantage of rescaled metrics in early identifying the significant patents, and to compare the obtained results with known results for scientific papers \citep{mariani2016identification}.

To quantify how well the different metrics recognize the significant patents short after their issuing, we focus on the $M_{20}= 77$ patents that are at least $20$ years old at the end of the dataset. By performing the evaluation procedure described in Section \ref{sec:age_evaluation}, we study how their average ranking ratio and identification rate depend on their age up to $20$ years after issuing. We focus thus on a fixed group of target patents, which allows us to gauge the impact of time on the metrics' performance\footnote{Patents less than are $10$ years old, for example, could not contribute to the age bins from $10$ to $20$ years after issuing. Were we including also them in the control group of significant patents, we would have ended up with a control group with different composition for different age bins, which would have confounded the temporal effects that we focus on here.}.

\subsubsection{Average ranking ratio}
In qualitative agreement with Fig.~\ref{rankings}, Fig.~\ref{short}A shows striking differences between the metrics' performance. Shortly after issuing, the rescaled metrics achieve an average ranking ratio much lower than that of the non-rescaled metrics. For example, one year after publication, PageRank's and rescaled PageRank's average ranking ratio are equal to $20.8$ and $1.6$, respectively, which indicates a performance advantage of one order of magnitude in favor of $R(p)$.
The gap between rescaled PageRank and PageRank (rescaled citation count and citation count) closes $12$ ($7$) years after issuing.
There is therefore a medium-term temporal window over which the rescaled metrics rank the significant patents remarkably better than the non-rescaled metrics.

Importantly, once we have suppressed the age bias of $c$ and $p$, we are able to reveal the advantage of using (higher-order) network information to rank the significant patents instead of simply counting citations, which manifests itself in the performance advantage of $R(p)$ over $R(c)$.

\subsubsection{Identification rate}
Fig.~\ref{short}B shows the dependence of the metrics' identification rate $f_{0.005}(\Delta t)$ as a function of patent age. This evaluation measure quantifies the fraction of significant patents ranked in the top $0.5\%$ by the metrics when they were $\Delta t$ years old.
The rescaled metrics outperform the non-rescaled metrics shortly after publication; the gap between rescaled and non-rescaled metrics closes 
eventually:
$p$'s performance reaches $R(p)$'s performance $12$ years after issuing, and $c$'s performance reaches $R(c)$'s performance $6$ years after issuing.
These two timescales are consistent with those observed for the average ranking ratio, and they define a temporal window over which the rescaled metrics achieve an improved identification of the significant patents.

\subsection{The role of the network structure}
\label{sec:dcm}
In this Section, we address the following question: to which extent can the improved performance of PageRank-related metrics be explained by citation count dynamics alone?
In other words, once we control for the effect of citation count dynamics and randomize the rest, can we reproduce the results in Fig.~\ref{short}?

To address these questions, we use the the Dynamic Configuration Model (DCM) introduced by \cite{ren2017time} to generate random networks that preserve the individual nodes' citation count time-series observed in the original network. Differently from the widely-used configuration model \citep{molloy1995critical}, the DCM preserves the original network's temporal linking patterns \citep{ren2017time}.
In the DCM, the total system time span $T$ is divided into $L$ layers of equal duration $\Delta t=T/L$. The randomized networks are thus generated by rewiring the existing connections, within each layer, by preserving each node's indegree and outdegree variation in that layer (see \cite{ren2017time} for the details). The expected number of edges $E_{ij}(n)$ from node $j$ to node $i$ at layer $n$ is given by
\begin{equation}
E_{ij}(n)=\frac{\Delta k_i^{in}(n)\,\Delta k_j^{out}(n)}{E(n)},
\end{equation}
where $\Delta k_i^{in}$ ($\Delta k_j^{out}(n)$) denotes the indegree (outdegree) increase of node $i$ ($j$) in layer $n$, and $E(n)$ denotes the total number of edges in layer $n$. In our work, we set $L=100$ which results in $\Delta t = 310$ days.

\begin{figure*}
\centering
\includegraphics[width=18cm]{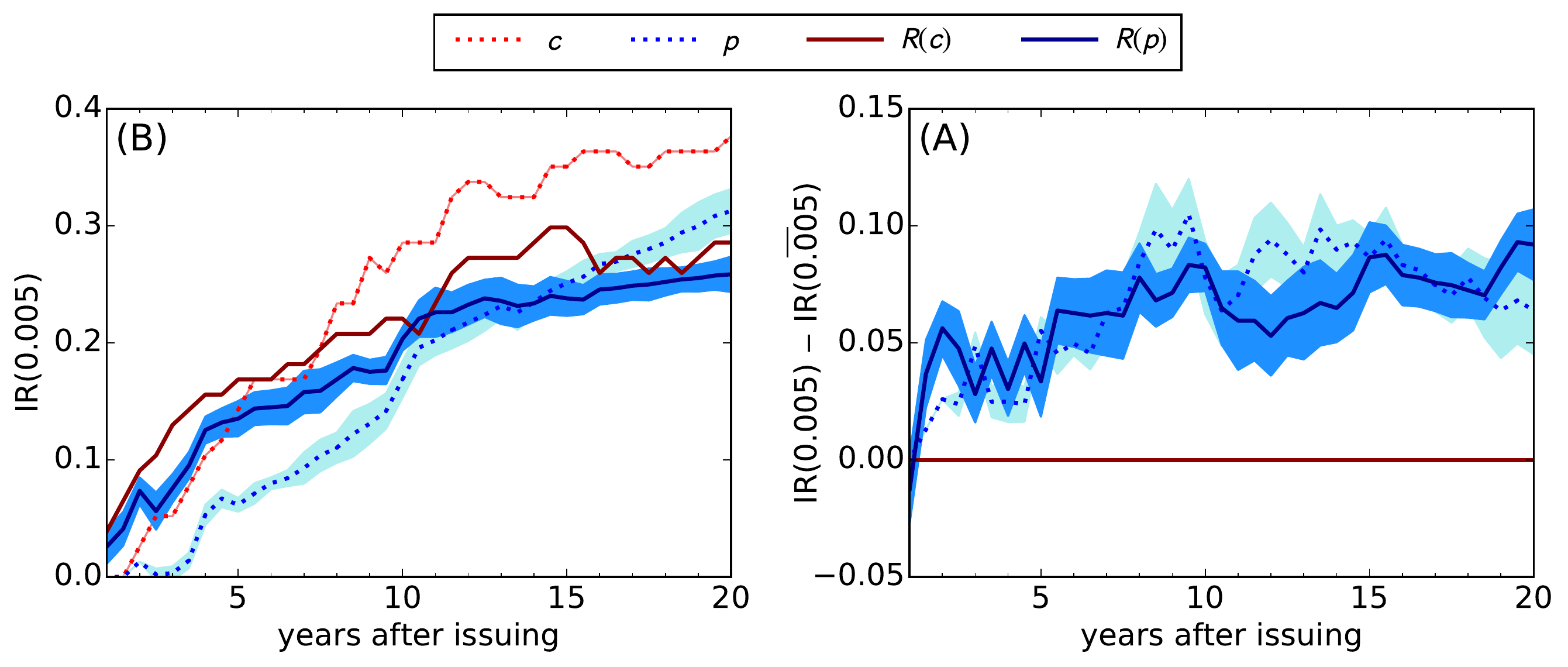}
\caption{\emph{(A)} Metrics' performance in identifying the significant patents in a random network generated with the Dynamic Configuration Model. The lines and the shaded areas around the lines represent the mean and the standard error, respectively, measured over $12$ realizations of the randomization process. \emph{(B)} Difference between the performance observed in the real data and that observed in the randomized networks generated with the Dynamic Configuration Model.}
\label{performance-rand}
\end{figure*}

We compare the metrics' performance in the thus-generated random networks with the performance observed in the real data.
By construction, the model preserves the indegree time-series of the original network; as a consequence, the performance of the citation count and rescaled citation count is the same as in the real data (Fig.~\ref{performance-rand}).
The model allows us to assess whether the advantage of network-based metrics (Fig.~\ref{performance-rand}) is a genuine network effect or if it can be explained by random fluctuations.

In the randomized networks, the network-based metrics have no advantage with respect to citation-counting metrics in identifying the significant patents (Fig.~\ref{performance-rand}A). In fact, $R(p)$ falls slightly below $R(c)$ for almost every patent age. Fig.~\ref{performance-rand}B shows that the performance difference between the performance of PageRank-related metrics in real and randomized networks is significantly positive. We conclude that controlling for the individual nodes' citation count dynamics is not sufficient to explain our findings. Therefore, (higher-order) network structure plays a significant role for the advantage of network-based metrics with respect to citation-counting metrics in identifying the significant patents.


\subsection{Top patents}
\label{sec:top_patents}
In this section we inspect the top-ranked patents. For simplicity, we focus on the top-$15$ patents as ranked by PageRank (Table~\ref{tab:pr}) and rescaled PageRank score (Table~\ref{tab:rpr}).

Table~\ref{tab:pr} shows that also patents with relatively few citations can reach the top of the ranking by PageRank score, which confirms the idea that in citation networks, the PageRank algorithm can identify ``hidden gems'' \citep{chen2007finding} that are underestimated by citation count. A paradigmatic example in this sense is patent $3813316$ (``Microorganisms having multiple compatible degradative energy-generating plasmids and preparation thereof''). The patent is ranked $6$th by PageRank despite having been cited only $16$ times. By inspecting the patent's neighborhood, it emerges that the reason for this is that the patent has been cited by patents with relatively large citation count and, additionally, small outdegree. For example, patent $3813316$ is the only patent cited by patent $4237224$ (``Process for producing biologically functional molecular chimeras'', $c=285$, included in the Strumsky-Lobo list of significant patents) which is ranked $3$rd by PageRank. Highly scoring patent $3813316$ refers only to patent $3723248$ (``Method for producing ketoglucaric acid'') which is consequently ranked $38$th by PageRank despite having received only one citation. Small outdegree of the citing patents is crucial because it implies that a large portion of the citing patents' score will contribute to the cited patent's score in the score redistribution process defined by Eq.~(\ref{pr}).

Table~\ref{tab:rpr} shows that the top-$15$ patents by rescaled PageRank span a wider temporal range (1934-2010) than the top-$15$ by PageRank (1942-1996), which is a direct consequence of the age-bias removal. 
On the other hand, Table~\ref{tab:rpr} also points out a potential limitation of the rescaling procedure. Among the $15$ top-ranked patents, four are indeed from $2010$ (the last year in the dataset) and received only one citation. This happens because only a few among the most recent patents received citations, which results in temporal windows with a large fraction of patents with zero citations. Within such a temporal window, a patent can achieve large rescaled score thanks to one single citation. A possible solution for this issue is to only include the patents whose temporal windows contain a certain minimal number of incoming citations. However, we prefer to show the scores of all the patents in order to highlight the subtleties associated with the evaluation of very recent patents.

\section{A comparison of the APS papers' and the US patents' citation network dynamics}
\label{sec:comparison}
Section~\ref{section:results} validates the rescaled metrics as better indicators of significance of recent patents than the non-rescaled metrics. Yet, there is a remarkable difference between the behavior of the identification rate observed in our analysis of the US patent dataset (Fig.~\ref{short}B) and that reported by \cite{mariani2016identification} in their analysis of the American Physical Society (APS) paper citation network: \cite{mariani2016identification} found that $R(p)$ ranks more than $30\%$ of the Physical Review Letters milestone letters in the top $0.5\%$ already one year after publication, whereas it only ranks $1 \%$ of the Strumsky-Lobo significant patents in the top $1\%$ one year after issuing.

The qualitative difference between Fig.~\ref{short}B and Fig.~3B in \cite{mariani2016identification} for significant papers motivated us to explore the differences between the dynamics of (significant) patents and that of (significant) papers. To this end, we analyzed an extension of the dataset\footnote{In particular, we analyzed the APS citation network from $1893$ to $2016$, which comprises $593,443$ papers and $7,031,030$ citations between them.} used by \cite{mariani2016identification}, and compared the obtained results with those obtained for the US patent citation network. The results of our analysis are summarized in Tables \ref{table:average}. The table shows that both the significant papers and the significant patents: (1) tend to be cited more than ordinary papers and ordinary patents, respectively, in the respective datasets; (2) tend to accrue citations faster than ordinary papers and ordinary patents, respectively.
Like patents of high economic value  \citep{lee2017makes}, the Strumsky-Lobo significant patents tend to receive the first few citations quicker than ordinary patents. 

\begin{table}[t]
\centering
\begin{tabular}{llrrr}
\toprule
Dataset & Group of nodes & citations & $\tau_3$ & $\tau_5$ \\
\midrule
 \textbf{US patents} & Significant patents & 105.6 & 9.6 y & 12.0 y \\
 & All patents & 7.4 & 24.9 y & 31.4 y \\
\midrule
 \textbf{APS papers} & Significant papers & 457.0 & 1.0 y & 1.4 y \\
 & All papers & 11.8 & 3.6 y & 4.8 y \\
\bottomrule
\end{tabular}
\caption{A comparison between the average properties of all nodes and the average properties of the significant nodes in the APS paper and US patent citation network. The significant nodes are the milestone letters and the Strumsky-Lobo significant patents in the two datasets, respectively.}
\label{table:average}
\end{table}

However, there is a striking difference between the dynamics of the two datasets: the APS papers tend to accrue citations much quicker than US patents. For example, the time needed for papers that received at least three total citations to receive their first three citations is much smaller for papers ($3.6$ years on average) than for patents ($24.9$ years). The same is true if we restrict the analysis to the significant papers ($1.0$ years) and patents ($9.6$ years), respectively.
These results point out that the smaller identification rate for patents shortly after issuing is partly a manifestation of the slower citation dynamics of patents with respect to the citation dynamics of papers.\footnote{
We emphasize that the APS papers dataset contains only citations among the APS papers. By considering the citations to APS papers from non-APS papers, $\tau_3$ and $\tau_5$ would further decrease and the difference from the patent dataset (which, by contrast to the APS dataset, is comparably complete) would further magnify.}

\section{Conclusions}

Our paper has two main messages.

First, we find that using the whole network topology instead of only counting citations brings a substantial advantage in identifying the significant patents. Both the observed degree-degree correlations (Fig.~\ref{degree_correlations}) and the performance edge of PageRank-related metrics over citation-counting metrics (Fig.~\ref{short}) suggest that important patents build on other other important patents. 
This supports the hypothesis that high-impact patents ``stand on the shoulders of giants'', in a similar way as scientific papers \citep{bornmann2010scientific}, although the high prevalence of examiner-added citations in patents makes the analogy imperfect.

Second, we show that removing the time bias of static centrality metrics allows one to identify significant patents much earlier than it is possible with conventional static metrics. 
The rescaling procedure which we use to remove the time bias is efficient and thus applicable even to large-scale datasets \citep{vaccario2017quantifying}.

There are some limitations to our work that deserve to be discussed. 
First of all, we have pointed out that the early-identification of significant patents is more difficult than that of significant papers, because patents take more time to accumulate citations (Section~\ref{sec:comparison}). Second, the time-rescaled metrics are based on the assumption that a good ranking of the patents should give the patents from different age periods the same chance to get to the top of the ranking. While this assumption is customary in paper citation analysis \citep{waltman2016review}, it creates a bias against patents that appear in periods of intensive breakthrough inventive activity, if they exist.
Third, the rescaled metrics evaluate the most recent patents on the basis of citations received in a relatively short time period. While this may be justified by the finding that patents in rapidly growing domains are highly cited shortly after issuing \citep{benson2015quantitative}, it potentially misses out ``sleeping-beauty'' \citep{ke2015defining} patents that received a substantial amount of citations only many years after issuing.

We see three major directions for extending this research.
The most obvious is to acknowledge that there are different citation practices across technological fields, just as different scientific fields exhibit different citation patterns \citep{waltman2016review}. Based on the results by \citet{vaccario2017quantifying}, we know that the rescaling procedure can in principle be extended to suppress the bias by technological field as well. However, while it is natural to suppress biases by scientific field of paper-level metrics due to their use in research evaluation, it remains unclear whether a similar approach would be the most effective strategy to rank patents. Besides, using patent classification information is problematic when the goal is to rigorously test predictive ability, because the classification system is changing often and many patents are reclassified \citep{lafond2017long}. Second, while there exist theoretical explanations for how the broad citation count distribution and the bias of citation-based metric by node age emerge as a result of the dynamics of the system \citep{valverde2007topology,newman2009first,medo2011temporal,mariani2015ranking}, a model-based explanation of the strong degree-degree correlations and the improved PageRank performance observed in our dataset is still lacking. 
Third, while we studied PageRank as a paradigmatic network-based metric because of its plausible assumption (``a node is important if it is cited by other important nodes''), other network-based metrics \citep{liao2017ranking} can be analyzed in a similar way to improve our understanding of which metrics best identify important patents.

\section*{Acknowledgments}
MSM and MM acknowledge financial support from the Swiss National Science Foundation Grant No. 200020-156188. FL acknowledges financial support from Partners for a New Economy and the Institute for New Economic Thinking.

\bibliographystyle{spbasic}      

\clearpage

\begin{table*}[t]
\caption{Top-15 patents as ranked by PageRank score $p$ (asterisks mark the Strumsky-Lobo significant patents).}
\label{tab:pr}
\begin{tabular}{rrp{9cm}rrr}
\toprule
Rank & Patent ID & Patent title & Issuing date & $c$ & $p \cdot 10^{5}$  \\
\midrule
$1$ & $4683195$ & Process for amplifying, detecting, and/or-cloning nucleic acid sequences & 28-7-1987 & $1956$ & $2.824$ \\
$2$ & $4683202$ & Process for amplifying nucleic acid sequences & 28-7-1987 & $2169$ & $ 2.691$\\
$3$ & $4237224$ & (*) Process for producing biologically functional molecular chimeras & 2-12-1980 & $285$ & $2.687$\\
$4$ & $4395486$ & Method for the direct analysis of sickle cell anemia   & 26-7-1983 & $71$ & $1.731$ \\
$5$ & $4723129$ & Bubble jet recording method and apparatus in which a heating element generates bubbles in a liquid flow path to project droplets   & 2-2-1988 & $1962$ & $1.416$\\
$6$ & $3813316$ & Microorganisms having multiple compatible degradative energy-generating plasmids and preparation thereof & 28-5-1974 & $16$ & $1.399$ \\
$7$ & $5536637$ & Method of screening for cDNA encoding novel secreted mammalian proteins in yeast   & 16-6-1996 & $422$ & $1.344$ \\
$8$ & $4558413$ & Software version management system & 10-12-1985 & $1956$ & $1.326$\\
$9$ & $4358535$ & Specific DNA probes in diagnostic microbiology  & 9-11-1982 & $436$ & $1.324$ \\
$10$ & $2297691$ & Electrophotography & 6-10-1942 & $588$ & $1.312$\\
$11$ & $4463359$ & Droplet generating method and apparatus thereof  & 31-7-1984  & $1694$ & $1.263$ \\
$12$ & $5523520$ & Mutant dwarfism gene of petunia & 4-6-1996 & $1139$ & $1.221$ \\
$13$ & $4812599$ & Inbred corn line PHV78 & 14-3-1989  & $179$ & $1.187$ \\
$14$ & $4740796$ & Bubble jet recording method and apparatus in which a heating element generates bubbles in multiple liquid flow paths to project droplets  & 26-4-1988  & $1663$ & $1.093$ \\
$15$ & $5103459$ & System and method for generating signal waveforms in a CDMA cellular telephone system  & 7-4-1992  & $1208$ & $1.034$ \\
\bottomrule
\end{tabular}
\end{table*}

\begin{table*}[t]
\caption{Top-15 patents as ranked by rescaled PageRank score $R(p)$ (with $\Delta_p=15,000$). Asterisks mark the Strumsky-Lobo significant patents.}
\label{tab:rpr}
\begin{tabular}{rrp{9cm}rrr}
\toprule
Rank & Patent ID & Patent title & Issuing date & $c$ & $R(p)$  \\
\midrule
$1$ & $7764447$ & Optical element holding device, lens barrel, exposing device, and device producing method & 27-7-2010  & $1$ & $104.9$ \\
$2$ & $4237224$ & (*) Process for producing biologically functional molecular chimeras & 2-12-1980  & $285$ & $99.1$ \\
$3$ & $2297691$ & Electrophotography & 6-10-1942  & $588$ & $91.8$ \\
$4$ & $7749477$ & Carbon nanotube arrays & 6-7-2010  & $1$ & $84.8$ \\
$5$ & $7784029$ & Network service for modularly constructing a software defined radio & 24-8-2010 & $1$ & $78.9$ \\
$6$ & $5536637$ & Method of screening for cDNA encoding novel secreted mammalian proteins in yeast  & 16-7-1996  & $422$ & $78.9$ \\
$7$ & $4683195$ & Process for amplifying, detecting, and/or-cloning nucleic acid sequences & 28-7-1987  & $1956$ & $78.2$ \\
$8$ & $5523520$ & Mutant dwarfism gene of petunia & 4-6-1996  & $1139$ & $76.9$ \\
$9$ & $4395486$ & Method for the direct analysis of sickle cell anemia   & 26-7-1983 & $71$ & $75.0$ \\
$10$ & $4683202$ & Process for amplifying nucleic acid sequences & 28-7-1987  & $2169$ & $74.6$ \\
$11$ & $7779788$ & Animal training system with multiple configurable correction settings & 24-8-2010  & $1$ & $73.3$ \\
$12$ & $1970578$ & Assistants for the textile and related industries & 21-8-1934  & $241$ & $73.3$ \\
$13$ & $3813316$ & Microorganisms having multiple compatible degradative energy-generating plasmids and preparation thereof & 28-5-1974 & $16$ & $72.7$ \\
$14$ & $5572643$ & Web browser with dynamic display of information objects during linking & 5-11-1996  & $1120$ & $71.8$ \\
$15$ & $4723129$ & Bubble jet recording method and apparatus in which a heating element generates bubbles in a liquid flow path to project droplets   & 2-2-1988 & $1962$ & $68.9$ \\
\bottomrule
\end{tabular}
\end{table*}

\end{document}